\documentclass[aps,pra,twocolumn,showpacs,superscriptaddress,floatfix]{revtex4-1}
\usepackage{graphicx,amsmath,amssymb}
\usepackage[usenames]{color}
\usepackage[dvipsnames]{xcolor}
\usepackage[unicode=true,pdfusetitle,
 bookmarks=false,
 breaklinks=false,pdfborder={0 0 1},backref=false,colorlinks,urlcolor=Black, linkcolor=blue,citecolor=blue]
 {hyperref}
\begin{document}

\title{Globalized Nonlinear Critical Quantum Metrology by Two-photon Rabi-Stark model
}
\author{Zu-Jian Ying}
\email{yingzj@lzu.edu.cn}
\affiliation{School of Physical Science and Technology, Lanzhou University, Lanzhou 730000, China}
\affiliation{Key Laboratory for Quantum Theory and Applications of MoE, Lanzhou Center for Theoretical Physics, Lanzhou University, Lanzhou 730000, China}

\begin{abstract}
Squeezing as a quantum resource for quantum metrology is robust against decoherence and dissipation, while the conventional nonlinear two-photon quantum Rabi model (QRM) provides a squeezing resource immune to the divergence problem of preparation time of probe state (PTPS). However the critical point of the two-photon QRM is locally restricted to one single point, which hinders a wider application. In the present work we propose to combine the Stark coupling with the two-photon QRM to realize a tunable critical point so that the nonlinear critical quantum metrology can be globalized. As demonstrated by the diverging quantum Fisher information (QFI) the protocol enables us to acquire a high measurement precision in a wide range of coupling parameter rather than locally at a single critical point. Moreover, We find that the QFI not only manifests criticality but also exhibits universality.  As a particular merit of our protocol, a strong squeezing can be globally retained as the leading quantum resource, while at the same time the PTPS remains in a finite order.

\end{abstract}
\pacs{ }
\maketitle


\section{Introduction}

Recently critical quantum metrology (CQM) based on the finite-component quantum phase
transitions in light-matter interactions has been arising as a promising application of quantum
resources in advanced quantum technologies~\cite{Garbe2020,Montenegro2021-Metrology,Chu2021-Metrology,Garbe2021-Metrology,Ilias2022-Metrology, Ying2022-Metrology,Gietka2023PRL-Squeezing,Hotter2024-Metrology,Alushi2024PRL,Mukhopadhyay2024PRL,Mihailescuy2024,
Ying-Topo-JC-nonHermitian-Fisher,Ying-g2hz-QFI-2024,Ying-g1g2hz-QFI-2025,Ying2025g2A4}.
In particular, nonlinear CQM by exploiting
effect of nonlinear coupling and criticality turns out to have more advantages~\cite{Ying2022-Metrology,Gietka2023PRL-Squeezing,Hotter2024-Metrology,Ying-g2hz-QFI-2024,Ying-g1g2hz-QFI-2025}.

Indeed, with the merits of high controllability and tunability, light-matter
interactions~\cite{Diaz2019RevModPhy,Kockum2019NRP,PRX-Xie-Anistropy,Eckle-Book-Models,JC-Larson2021,Boite2020,Qin-ExpLightMatter-2018,WangYouJQ2023DeepStrong,Qin2024PhysRep,LiPengBo-Magnon-PRL-2024}
are building an ideal platform with great potential for applications in CQM~%
\cite{Garbe2020,Garbe2021-Metrology,Ilias2022-Metrology,Ying2022-Metrology,Hotter2024-Metrology,Ying-Topo-JC-nonHermitian-Fisher}. A fascinating phenomenon in the continuous coupling enhancements of light-matter
interactions is the existence of quantum phase transitions (QPTs)~\cite{Liu2021AQT,Ashhab2013,Ying2015,Hwang2015PRL,Hwang2016PRL,Irish2017,
Ying-g2hz-QFI-2024,Ying-g1g2hz-QFI-2025,Ying2025g2A4,
LiuM2017PRL,Ying-2018-arxiv,Ying2020-nonlinear-bias,Ying-2021-AQT,
Ying-gapped-top,Ying-Stark-top,Ying-Spin-Winding,
Ying-JC-winding,Ying-Topo-JC-nonHermitian,Ying-Topo-JC-nonHermitian-Fisher,Ying-gC-by-QFI-2024, Grimaudo2022q2QPT,Grimaudo2023-Entropy,Grimaudo2024PRR,Zhu2024PRL,DeepStrong-JC-Huang-2024,PengJie2019,Padilla2022,Gao2022Rabi-dimer,GaoXL2025SPT}. Actually a most fundamental model of light-matter
interactions is the quantum Rabi model (QRM)~\cite{rabi1936,Braak2011,Rabi-Braak,Eckle-Book-Models}. The QRM has a finite-component QPT, with the low-frequency limit as a replacement of the thermodynamical limit in condense matter~\cite{Ashhab2013,Ying2015,Hwang2015PRL,LiuM2017PRL,Irish2017}, which has been firstly applied for CQM~\cite{Garbe2020,Garbe2021-Metrology}. More abundant QPTs emerge by introducing anisotropy, bias and nonlinear couplings~\cite{
LiuM2017PRL,Ying-2018-arxiv,Ying2020-nonlinear-bias,Ying-2021-AQT,
Ying-gapped-top,Ying-Stark-top,Ying-Spin-Winding,
Ying-JC-winding,Ying-Topo-JC-nonHermitian,Ying-Topo-JC-nonHermitian-Fisher,Ying-gC-by-QFI-2024}, including Landau-class and topological-class transitions~\cite{
Ying-2021-AQT,Ying-gapped-top,Ying-Stark-top}, coexistence and even simultaneous occurrence of these two classes of transitions~\cite{
Ying-2021-AQT,Ying-Stark-top,Ying-JC-winding,Ying-Topo-JC-nonHermitian-Fisher}, topological transitions conventionally with and unconventionally without gap closing~\cite{
LiuM2017PRL,Ying-2018-arxiv,Ying2020-nonlinear-bias,Ying-2021-AQT,
Ying-gapped-top,Ying-Spin-Winding,
Ying-JC-winding}, transition of amplitude squeezing and phase squeezing~\cite{Ying-gapped-top} and so forth.
This mini-world of QPTs~\cite{Ying-gapped-top} provides various quantum resources for potential applications in CQM~\cite{Ying-g2hz-QFI-2024,Ying-g1g2hz-QFI-2025}.

In the protocol developments of CQM, nonlinear CQM manifests some more merits~\cite{Ying2022-Metrology,Gietka2023PRL-Squeezing,Hotter2024-Metrology,Ying-g2hz-QFI-2024,Ying-g1g2hz-QFI-2025}. Actually
CQM based on the linear QRM has been confronted with restriction to low frequency limit and the divergence problem of preparation time of probe state (PTPS)~\cite{Garbe2020,Garbe2021-Metrology,Ying2022-Metrology,Gietka2022-ProbeTime}. Nevertheless, these problems can be resolved in the nonlinear CQM in terms of quadratic nonlinear coupling which is applicable to finite frequencies and yields finite PTPS~\cite{Ying2022-Metrology,Ying-g2hz-QFI-2024,Ying-g1g2hz-QFI-2025}. The conventional model for the quadratic nonlinear coupling is the two-photon QRM~\cite{Felicetti2018-mixed-TPP-SPP,Felicetti2015-TwoPhotonProcess,e-collpase-Garbe-2017,Rico2020,e-collpase-Duan-2016,CongLei2019}.
The two-photon QRM can give rise to squeezing which is a primary quantum resource most widely explored for quantum metrology~\cite{Maccone2020Squeezing,Lawrie2019Squeezing,Gietka2023PRL-Squeezing,Gietka2023PRL2-Squeezing,Candeloro2021-Squeezing}. Note that, in contrast to the vulnerability of entanglement as quantum resource~\cite{Horodecki2009entanglement}, squeezing is robust against decoherence and dissipation~\cite{Gietka2023PRL-Squeezing,LiJieYouJQ2022SqueezingAgaistTemperature,WengYouJQ2025SqueezingRobust,Buzek1992SqueezingAgainstdissipation}. However, a strong squeezing is available in the two-photon QRM only for a coupling around the critical point $g_{\rm T}$, while it is much weakened away from $g_{\rm T}$~\cite{Ying-g2hz-QFI-2024,Ying-g1g2hz-QFI-2025}. It would be favorable to have a globally strong squeezing as the quantum resource in nonlinear CQM.

In the present work we propose to combine the Stark coupling with
the conventional quadratic coupling to globalize the nonlinear CQM.
The protocol can be realized by the two-photon Rabi-Stark model in
light-matter interaction. The introduction of the Stark coupling extend the
single critical point $g_{\mathrm{T}}$ in the conventional two-photon QRM to a continuously
tunable critical point, thus the local restriction
of the critical behavior can be broken and the criticality is available in a
wide range of coupling for application of CQM, as demonstrated by divergent quantum Fisher information (QFI).
We extract the critical exponent of QFI and find that the critical QFI manifests universality over different couplings.
Such a globalized nonlinear CQM not only keeps the merits of applicability to finite frequencies
and maintaining a finite PTPS, but also can retain a globally strong squeezing.

The paper is organized as follows.
Section~\ref{Section-Model}
introduces the two-photon Rabi-Stark model and gives the critical point in the absence and the tunable critical point in the presence of the Stark coupling.
Section~\ref{Section-QFI}
addresses the globally divergent quantum Fisher information (QFI) for CQM, with sensitivity resources analyzed and critical exponent obtained.
Section~\ref{Section-Global-Squeezing}
demonstrates the globally strong squeezing by the Wigner function.
Section~\ref{Section-Finite-T}
shows the finite PTPS.
Finally, Section~\ref{Section-Conclusion}
gives a summary of conclusions.

\section{Two-photon Rabi-Stark model}
\label{Section-Model}

We consider both two-photon coupling and Stark coupling \cite%
{Felicetti2018-mixed-TPP-SPP,Stark-Grimsmo2013,Eckle-2017JPA,Stark-Cong2020,Ying-Stark-top,ChenQH2020TwoPhotonStark,Zhai2025TwoPhotonStark}
described generally here by the following Hamiltonian%
\begin{equation}
H=\omega a^{\dagger }a+\frac{\Omega }{2}\hat{\sigma}_{x}+g_{2}\hat{\sigma}%
_{z}[a^{\dagger 2}+a^{2}+\chi _{z}(2\hat{n}+1)]+\omega \chi \hat{n}\hat{%
\sigma}_{x}  \label{H-g2-Stark}
\end{equation}%
which can be realized in superconducting circuits or trapped ions\cite%
{Felicetti2018-mixed-TPP-SPP,Stark-Grimsmo2013,Stark-Cong2020,Zhai2025TwoPhotonStark}%
. The model describes coupling between a bosonic mode, with frequency $%
\omega $ and photon number $\hat{n}=a^{\dagger }a$ created (annihilated) by $%
a^{\dagger }$ ($a),$ and a qubit represented by the Pauli matrices $\hat{%
\sigma}_{x,y,z}$. Here $\Omega $ denotes the level splitting in a cavity
system and the tunneling energy of flux qubit in a superconducting circuit
system~\cite{flux-qubit-Mooij-1999}. In the standard QRM the coupling is
linear, involving a one-photon process of emission and absorption denoted by
$g_{1}\hat{\sigma}_{z}(a^{\dagger }+a)$ \cite%
{rabi1936,Braak2011,Rabi-Braak,Eckle-Book-Models}. The coupling in our
Hamiltonian (\ref{H-g2-Stark}) is nonlinear, including both the quadratic
coupling ($g_{2}$ term)\cite{Ying-2018-arxiv,Ying2020-nonlinear-bias} and
the Stark ($\chi $ term) coupling \cite%
{Eckle-2017JPA,Stark-Cong2020,Ying-Stark-top}. Here we have set a more
general case for the quadratic coupling by introducing the parameter $\chi
_{z}$. Setting $\chi _{z}=\chi =0$ retrieves the two-photon QRM~\cite%
{Felicetti2018-mixed-TPP-SPP,Felicetti2015-TwoPhotonProcess,e-collpase-Garbe-2017,Rico2020,e-collpase-Duan-2016,CongLei2019}%
, while $\chi _{z}=1$ gives rise to the full form of quadratic coupling $%
g_{2}\hat{\sigma}_{z}(a^{\dagger }+a)^{2}$ which is more originally in
superconducting circuits~\cite{Felicetti2018-mixed-TPP-SPP}.

The standard linear QRM has the $Z_{2}$ symmetry $P_{2}=\sigma _{x}e^{i\pi
a^{\dagger }a}$. Here, in the nonlinear couplings, the $\chi _{z}=0$ case
has the $Z_{4}$ symmetry $P_{4}=\sigma _{x}e^{i\pi a^{\dagger }a/2}$ which
is broken in the $\chi _{z}=1$ case and replaced by another $Z_{2}$ symmetry
$P_{x}=e^{i\pi a^{\dagger }a}$~\cite%
{Ying-g2hz-QFI-2024,Ying-2021-AQT,Ying-JC-winding}.

\begin{figure}[t]
\centering
\includegraphics[width=1.0\columnwidth]{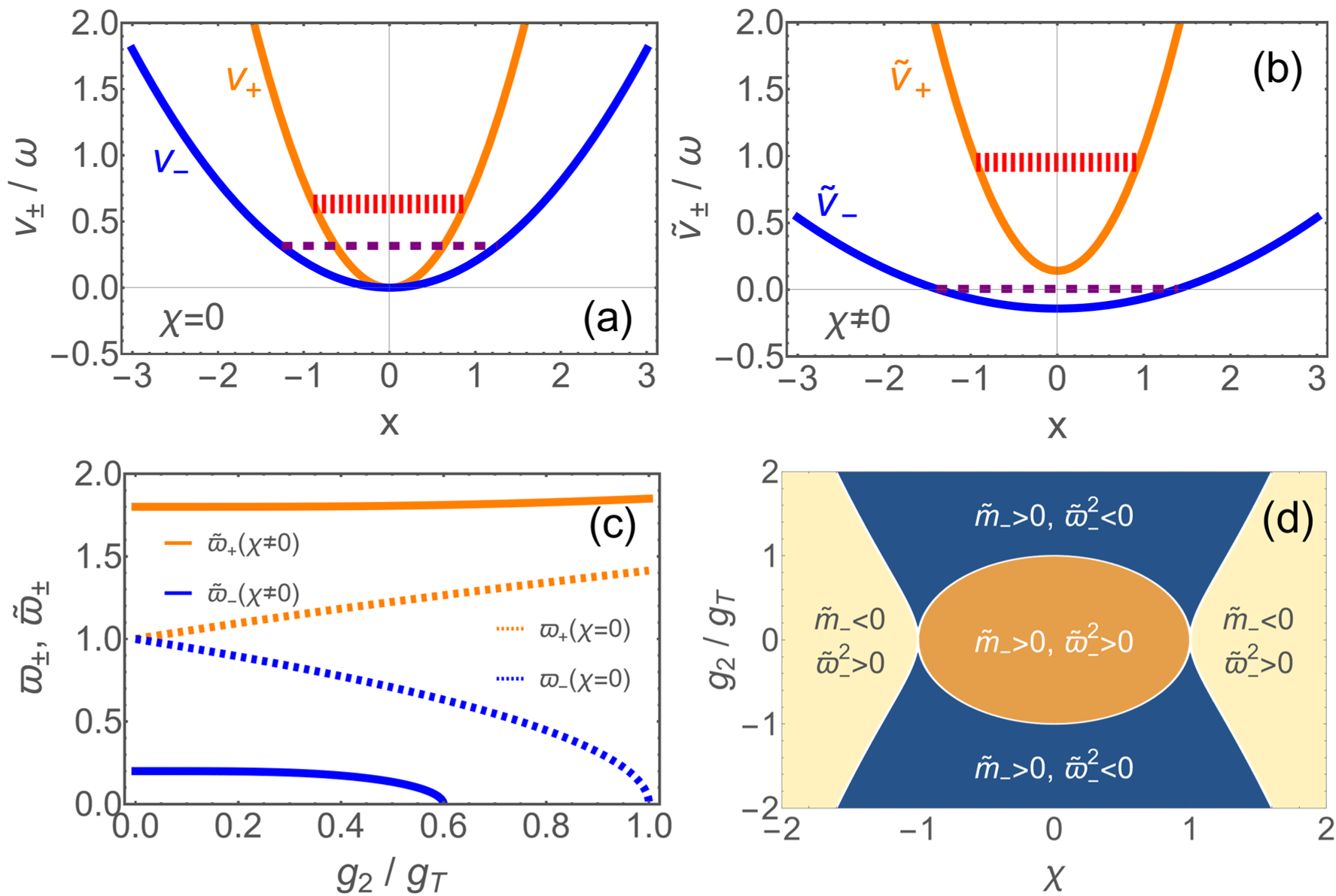}
\caption{Tunable critical point by Stark coupling $\chi$.
(a) Effective potential $v_{\pm}$ at $\chi =0$.
(b) Wider potential $\tilde{v}_{-}$ and narrower $\tilde{v}_{+}$ at $\chi \neq 0$.
(c) Potential frequency at $\chi =0$ ($\varpi _{\pm}$, broken lines) and $\chi \neq 0$ ($\tilde{\varpi} _{\pm}$, solid lines).
(d) Phase diagram of $\tilde{m} _{-}$ and $\tilde{\varpi} _{-}$ in the $\chi$-$g_2$ plane. The dotted and dashed lines in (a,b) represent the effective spin-dependent single-particle energy. In (a-c) $\chi=0.8$ is illustrated for $\chi \neq 0$ case.}
\label{fig-v-m-w}
\end{figure}

\section{Spin-rotated potential and tunable critical point}

\label{Section-Tunable-gC}

As the quadratic coupling and the Stark coupling are nontrivially involving
the photons and different spin directions, we shall map to the position
space and introduce a spin rotation to facilitate the analysis of physical
picture and the extracting of the critical point.

\subsection{In the absence of the Stark coupling}

By the quadrature transformation $a^{\dagger }=(\hat{x}-i\hat{p})/\sqrt{2}$,
$a=(\hat{x}+i\hat{p})/\sqrt{2}$ with position $x$ and momentum $\hat{p}=-i%
\frac{\partial }{\partial x}$, we can rewrite $H$ in the position space as $%
H_{x}$. In the absence of the Stark coupling, the Hamiltonian becomes
\begin{eqnarray}
H_{x} &=&\frac{\omega }{2}\hat{p}^{2}+\frac{\omega }{2}\left( 1+\overline{g}%
_{2}\hat{\sigma}_{z}\right) \hat{x}^{2}+\frac{\Omega }{2}\hat{\sigma}%
_{x}-w_{z}\frac{\omega \overline{g}_{2}}{2}\hat{\sigma}_{z}\hat{p}^{2} \\
&=&\omega \left( \frac{\hat{p}^{2}}{2m_{\sigma }}+\frac{1}{2}m_{\sigma
}\varpi _{\sigma }^{2}x^{2}\right) +\frac{\Omega }{2}\hat{\sigma}_{x}
\end{eqnarray}%
up to a constant energy $E_{0}=-\omega /2$, where $m_{\sigma }$ and $\varpi
_{\sigma }$ are spin-dependent effective mass and frequency
\begin{eqnarray}
m_{\sigma } &=&\left( 1-w_{z}\overline{g}_{2}\hat{\sigma}_{z}\right) ^{-1},
\\
\varpi _{\sigma }^{2} &=&\left( 1+\overline{g}_{2}\hat{\widetilde{\sigma }}%
_{z}\right) \left( 1-w_{z}\overline{g}_{2}\hat{\sigma}_{z}\right) .
\end{eqnarray}%
Here we have defined $w_{z}=(1-\chi _{z})/(1+\chi _{z})$ and $\overline{g}%
_{2}=g_{2}/g_{\mathrm{T}}$ where
\begin{equation}
g_{\mathrm{T}}=\frac{\omega }{2\left( 1+\chi _{z}\right) }.
\end{equation}%
So the conventional two-photon coupling and full form of quadratic coupling
can be rescaled by the factor $\left( 1+\chi _{z}\right) $. Still, for the
two-photon QRM with $\chi _{z}=0$, the single-particle energy is degenerate
as the potential frequency becomes spin-independent $\varpi _{\pm }=\sqrt{1-%
\overline{g}_{2}^{2}}$, while the degeneracy is lifted for $\chi _{z}=1$ in
the full quadratic coupling as $\varpi _{\pm }=\sqrt{1\pm \overline{g}_{2}}$%
. The frequency difference actually creates chances for non-monotonous
degeneracy lifting in resource-combined quantum metrology\cite{Ying-g2hz-QFI-2024}. Nevertheless,
despite of the frequency difference, in both case we have the same form of
potential
\begin{equation}
v_{\pm }\left( x\right) =\frac{\omega }{2}m_{\sigma }\varpi _{\sigma
}^{2}x^{2}=\frac{\omega }{2}\left( 1\pm \overline{g}_{2}\right) \hat{x}^{2}
\end{equation}%
which differs for the two spin components. In this work we shall illustrate
by $\chi _{z}=1$ in figures. We show an example of $v_{\pm }\left( x\right) $
in Fig. \ref{fig-v-m-w}(a) where one can see that $v_{+}\left( x\right) $
(orange) is narrower while $v_{-}\left( x\right) $ (blue) is wider due to
the renormalization factor $m_{\sigma }\varpi _{\sigma }^{2}=\left( 1\pm
\overline{g}_{2}\right) $.

In particular, as shown by the effective potential frequency $\varpi _{-}$
[blue (dark gray) dotted line] in Fig. \ref{fig-v-m-w}(c) of the spin-down
component, there is a critical point
\begin{equation}
\overline{g}_{2c}=\frac{g_{2c}}{g_{\mathrm{T}}}=1
\end{equation}%
where the potential becomes flat and the spectral collapse occurs\cite%
{Felicetti2018-mixed-TPP-SPP,Felicetti2015-TwoPhotonProcess,e-collpase-Garbe-2017,Rico2020,e-collpase-Duan-2016,CongLei2019}%
. Above the critical point the harmonic potential turns to be downward, as a
consequence the system is unstable due to unbounded lower energy\cite%
{CongLei2019}.

\begin{figure}[t]
\centering
\includegraphics[width=1.0\columnwidth]{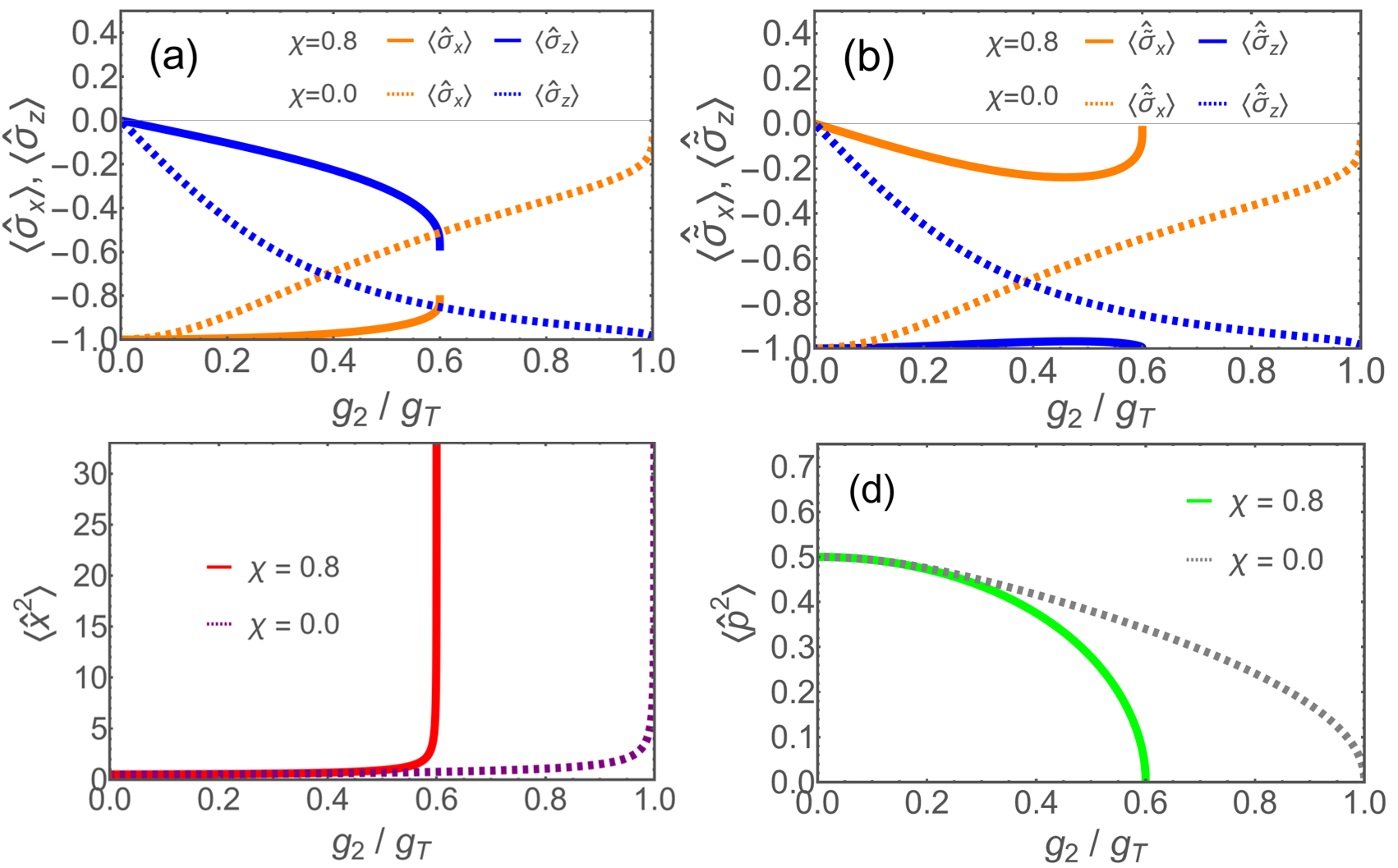}
\caption{Criticality both at zero and finite $\chi$.
(a) Spin expectation $\langle\hat{\sigma}_{x,z}\rangle$.
(b) Rotated spin expectation $\langle\hat{\tilde{\sigma}}_{x,z}\rangle$.
(c) $\langle\hat{x}^2\rangle$.
(d) $\langle\hat{p}^2\rangle$. Here broken (solid) lines denote the $\chi =0$ ($\chi \neq 0$) case in all panels. Here the results are obtained by exact diagonalization (ED).}
\label{fig-x2-p2-spinxz}
\end{figure}

Before $\overline{g}_{2c}$ the ground state properties are critical, as
shown by the diverging $\langle \widehat{x}^{2}\rangle $ and vanishing $%
\langle \widehat{p}^{2}\rangle $ in Figs. \ref{fig-x2-p2-spinxz}(c) and \ref%
{fig-x2-p2-spinxz}(d). Such a critical behavior can be exploited for
CQM. However, the critical point is restricted to a single point,
which hinders a wider application. In the following we introduce the Stark
coupling to break such a restriction.

\subsection{In the presence of the Stark coupling}

In the presence of the Stark coupling, the Hamiltonian in the position space
reads%
\begin{eqnarray}
H_{x} &=&\frac{\omega }{2}\hat{p}^{2}+\frac{\omega }{2}\left( 1+\overline{g}%
_{2}\hat{\sigma}_{z}+\chi \hat{\sigma}_{x}\right) \hat{x}^{2}+\frac{\Omega
_{\chi }}{2}\hat{\sigma}_{x}  \nonumber \\
&&+\frac{\omega }{2}\left[ \chi \hat{\sigma}_{x}-w_{z}\overline{g}_{2}\hat{%
\sigma}_{z}\right] \hat{p}^{2}  \label{Hx-Stark}
\end{eqnarray}%
where $\Omega _{\chi }=\Omega -\chi \omega $. As the $\hat{x}^{2}$ term
involves different spin directions, we further introduce a spin rotation in
the $\sigma _{x}$-$\sigma _{z}$ plane
\begin{eqnarray}
\hat{\widetilde{\sigma }}_{z} &=&\frac{\overline{g}_{2}\hat{\sigma}_{z}+\chi
\hat{\sigma}_{x}}{\sqrt{\overline{g}_{2}^{2}+\chi ^{2}}},\qquad \hat{%
\widetilde{\sigma }}_{x}=\frac{-\chi \hat{\sigma}_{z}+\overline{g}_{2}\hat{%
\sigma}_{x}}{\sqrt{\overline{g}_{2}^{2}+\chi ^{2}}}, \label{Sxz-tilde}\\
\hat{\sigma}_{z} &=&\frac{\overline{g}_{2}\hat{\widetilde{\sigma }}_{z}-\chi
\hat{\widetilde{\sigma }}_{x}}{\sqrt{\overline{g}_{2}^{2}+\chi ^{2}}},\qquad
\hat{\sigma}_{x}=\frac{\chi \hat{\widetilde{\sigma }}_{z}+\overline{g}_{2}%
\hat{\widetilde{\sigma }}_{x}}{\sqrt{\overline{g}_{2}^{2}+\chi ^{2}}},
\end{eqnarray}%
so that the Hamiltonian is reformed to be
\begin{equation}
\widetilde{H}_{x}=\omega \left( \frac{\hat{p}^{2}}{2\widetilde{m}_{\sigma }}+%
\frac{1}{2}\widetilde{m}_{\sigma }\widetilde{\varpi }_{\sigma
}^{2}x^{2}\right) +\frac{\widetilde{\Omega }}{2}\hat{\widetilde{\sigma }}%
_{x}+\widetilde{\epsilon }\hat{\widetilde{\sigma }}_{z}+\widetilde{\kappa }%
_{2}\hat{p}^{2}\hat{\widetilde{\sigma }}_{x}  \label{Hx-Stark-rotated}
\end{equation}
which is composed of coupled harmonic oscillators with reformed
spin-dependent mass and potential frequency:
\begin{eqnarray}
\widetilde{m}_{\sigma } &=&\left[ 1+\left( \chi ^{2}-w_{z}\overline{g}%
_{2}^{2}\right) \frac{\ \hat{\widetilde{\sigma }}_{z}}{\sqrt{\overline{g}%
_{2}^{2}+\chi ^{2}}}\right] ^{-1}, \label{mass-tilde} \\
\widetilde{\varpi }_{\sigma }^{2} &=&\left( 1+\sqrt{\overline{g}%
_{2}^{2}+\chi ^{2}}\hat{\widetilde{\sigma }}_{z}\right) \left[ 1+\frac{%
\left( \chi ^{2}-w_{z}\overline{g}_{2}^{2}\right) \ \hat{\widetilde{\sigma }}%
_{z}}{\sqrt{\overline{g}_{2}^{2}+\chi ^{2}}}\right] .
\end{eqnarray}%
The new spin basis for $\widetilde{\sigma }_x$ is given in Appendix \ref{Appendix-Spin-Rotation}.

Now we have renormalized spin-flipping strength $\widetilde{\Omega }$,
induced bias $\widetilde{\epsilon }$ and nonlinear spin-orbit coupling $%
\widetilde{\kappa }_{2}$%
\begin{eqnarray}
\widetilde{\Omega } &=&\frac{\Omega _{\chi }\overline{g}_{2}}{\sqrt{%
\overline{g}_{2}^{2}+\chi ^{2}}}, \label{Omega-tilde} \\
\widetilde{\epsilon } &=&\frac{\chi \Omega _{\chi }}{2\sqrt{\overline{g}%
_{2}^{2}+\chi ^{2}}}, \label{epsilon-tilde} \\
\widetilde{\kappa } &=&\frac{\omega }{2}\frac{\left( 1+w_{z}\right) \chi
\overline{g}_{2}}{\sqrt{\overline{g}_{2}^{2}+\chi ^{2}}}. \label{kapa-tilde}
\end{eqnarray}%
The effective potential is modified from $v_{\pm }\left( x\right) $ to be
\begin{equation}
\widetilde{v}_{\pm }\left( x\right) =\frac{\omega }{2}\widetilde{m}_{\sigma }%
\widetilde{\varpi }_{\sigma }^{2}x^{2}=\frac{\omega }{2}\left( 1\pm \sqrt{%
\overline{g}_{2}^{2}+\chi ^{2}}\right) \hat{x}^{2}.
\end{equation}%
We show $\widetilde{v}_{\pm }\left( x\right) $ in the presence of the Stark
coupling in Fig. \ref{fig-v-m-w}(b), with the same $\overline{g}_{2}$ as
Fig. \ref{fig-v-m-w}(a) in the absence of the Stark coupling. We see\ a
wider $\widetilde{v}_{-}\left( x\right) $ is yielded than $v_{-}\left(
x\right) $,\ which indicates that $\widetilde{v}_{-}\left( x\right) $ can
reach the flat potential earlier, as more clearly demonstrated by the $%
\widetilde{\varpi }_{-}$ in Fig. \ref{fig-v-m-w}(c). Indeed, the presence of
the Stark coupling leads to a reduced critical point
\begin{equation}
\overline{g}_{2c}^{\chi }=\sqrt{1-\chi ^{2}}.  \label{gC-at-Stark}
\end{equation}%
The phase diagram of $\widetilde{m}_{\sigma }$ and $\widetilde{\varpi }%
_{\sigma }$ is presented in Fig. \ref{fig-v-m-w}(d) where the system is
stable in the round region due to positive mass and real frequency. The
round boundary denotes the new $\chi $-dependent critical point $\overline{g}%
_{2c}^{\chi }$.

Note now that the critical point $\overline{g}_{2c}$ is not any longer
restricted to the single point $\overline{g}_{2c}=g_{2c}/g_{\mathrm{T}}=1$,
instead it can be continuously tuned by the Stark coupling, thus covering
the overall range of coupling in $g_{2}\in \lbrack 0,g_{\mathrm{T}}]$.

\subsection{Physical properties and criticality both in the absence and
presence of the Stark coupling}

In the presence of the Stark coupling, the criticality is retained and even
becomes more obvious, as shown by the ground-state spin expectations on
unrotated basis ($\langle \hat{\sigma}_{z,x}\rangle $) and rotated basis ($%
\langle \hat{\widetilde{\sigma }}_{z,x}\rangle $) as well as $\langle
\widehat{x}^{2}\rangle $ and $\langle \widehat{p}^{2}\rangle $ (solid lines)
in Figs.\ref{fig-x2-p2-spinxz}(a)-\ref{fig-x2-p2-spinxz}(d).

We first look at the spin expectations. From the reversed blue and orange
lines in Figs.\ref{fig-x2-p2-spinxz}(a) and \ref{fig-x2-p2-spinxz}(b), we
find that $\langle \hat{\sigma}_{z,x}\rangle $ and $\langle \hat{\widetilde{%
\sigma }}_{z,x}\rangle $ are reversed with respect to the $x,z$ components
at the $\overline{g}_{2}=0$ limit. Indeed in this limit, in the unrotated
spin representation the Hamiltonian (\ref{Hx-Stark}) becomes
\begin{eqnarray}
H_{x} &\rightarrow &\frac{\omega }{2}\hat{p}^{2}+\frac{\omega }{2}\left(
1+\chi \hat{\sigma}_{x}\right) \hat{x}^{2}+\frac{\Omega _{\chi }}{2}\hat{%
\sigma}_{x}  \nonumber \\
&&+\frac{\omega }{2}\chi \hat{\sigma}_{x}\hat{p}^{2}
\end{eqnarray}%
which is degenerate with respect to $\sigma _{z}$, thus $\langle \hat{\sigma}%
_{z}\rangle \rightarrow 0$. On the other hand, the degenerate $\sigma _{z}$
basis also allows a maximum spin flipping, leading to $\langle \hat{\sigma}%
_{x}\rangle \rightarrow -1$. In contrast, in the rotated spin
representation, the Hamiltonian (\ref{Hx-Stark-rotated}) has a finite
effective bias $\widetilde{\epsilon }$ and no spin flipping
\begin{equation}
\widetilde{H}_{x}\rightarrow \omega \left( \frac{\hat{p}^{2}}{2\widetilde{m}%
_{\sigma }}+\frac{1}{2}\widetilde{m}_{\sigma }\widetilde{\varpi }_{\sigma
}^{2}x^{2}\right) +\widetilde{\epsilon }\hat{\widetilde{\sigma }}_{z},
\end{equation}%
so that the spin is fully polarized $\langle \hat{\widetilde{\sigma }}%
_{z}\rangle \rightarrow -1$ but $\langle \hat{\widetilde{\sigma }}%
_{x}\rangle \rightarrow 0$. As a consequence, $\langle \hat{\sigma}%
_{z,x}\rangle $ and $\langle \hat{\widetilde{\sigma }}_{z,x}\rangle $ turn
out to be reversed.

Around the critical point, both $\langle \hat{\sigma}_{z}\rangle $ and $%
\langle \hat{\sigma}_{x}\rangle $ are finite while $\langle \hat{\widetilde{%
\sigma }}_{x}\rangle $ tends to be vanishing in saturated $\langle \hat{%
\widetilde{\sigma }}_{z}\rangle $. Although both $\langle \hat{\sigma}%
_{z,x}\rangle $ and $\langle \hat{\widetilde{\sigma }}_{z,x}\rangle $
manifest more or less critical behavior, the situation of $\langle \hat{%
\widetilde{\sigma }}_{z,x}\rangle $ is more convenient to see the leading
contributions in the QFI analyzed in next section.

The criticality can be more clearly seen from $\langle \widehat{x}%
^{2}\rangle $ and $\langle \widehat{p}^{2}\rangle $ (solid lines) in Figs.%
\ref{fig-x2-p2-spinxz}(c) and \ref{fig-x2-p2-spinxz}(d). Similarly to the
case in the absence of the Stark coupling, $\langle \widehat{x}^{2}\rangle $
is diverging and $\langle \widehat{p}^{2}\rangle $ is dropping, irrespective
of the spin basis rotation.

Besides the above facilitated extracting of the critical point, we will see
that the rotated basis also is more convenient for sensitivity resource
analysis for the QFI as in Sections \ref{Sect-resource-tracking} and \ref{Sect-exponent}.

\begin{figure}[t]
\centering
\includegraphics[width=1.0\columnwidth]{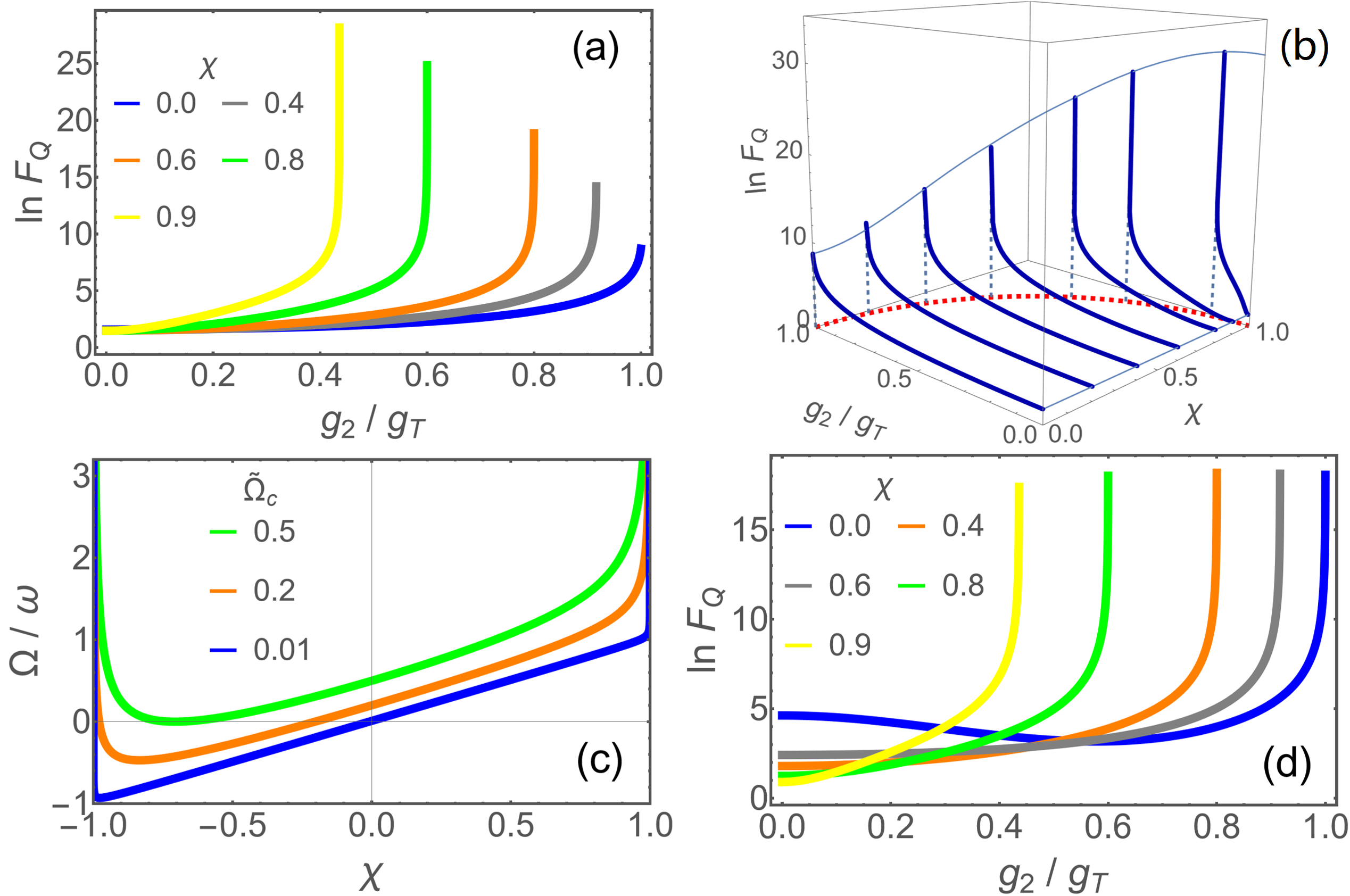}
\caption{Diverging quantum Fisher information (QFI) $F_Q$ by ED.
(a) $F_q$ (in natural logarithm) at different values of $\chi$ at fixed $\Omega=1.0\omega$.
(b) 3D plot for $F_Q$ vs $g_2$ and $\chi$ at fixed $\Omega=\omega$.
(c) Equi-$\tilde\Omega _c$ lines in the $\chi$-$\Omega$ plane.
(d) $F_Q$ at different values of $\chi$ with fixed $\tilde\Omega _c=0.2\omega$. We set $\omega=1$ as the unit throughout all figures.}
\label{fig-Fq}
\end{figure}

\section{Globalized nonlinear critical quantum metrology}

\label{Section-Globlized-QM}

\subsection{Quantum Fisher Information (QFI) for quantum metrology}

\label{Section-QFI}

In quantum metrology the measurement precision of experimental estimation on
a parameter $\lambda $ is bounded by $F_{Q}^{1/2}$ as limited by the quantum
Cram\'{e}r-Rao theorem~\cite{Cramer-Rao-bound}. Here $F_{Q}$ is the QFI
defined as \cite{Cramer-Rao-bound,Taddei2013FisherInfo,RamsPRX2018}
\begin{equation}
F_{Q}\left( \lambda \right) =4\left[ \langle \psi ^{\prime }\left( \lambda
\right) |\psi ^{\prime }\left( \lambda \right) \rangle -\left\vert \langle
\psi ^{\prime }\left( \lambda \right) |\psi \left( \lambda \right) \rangle
\right\vert ^{2}\right]   \label{Fq}
\end{equation}%
for a pure state $|\psi (\lambda )\rangle $. The prime symbol $^{\prime }$
denotes the derivative with respect to the parameter $\lambda $. A larger
QFI would mean a higher measurement precision. It has been shown that for a
real wave function $\psi (\lambda )$ the QFI can be simplified to be~\cite%
{Ying-gC-by-QFI-2024}
\begin{equation}
F_{Q}=4\langle \psi ^{\prime }\left( \lambda \right) |\psi ^{\prime }\left(
\lambda \right) \rangle .
\end{equation}%
Such a case usually occurs in non-degenerate states of a real Hamiltonian,
which also applies for the ground state of our model (\ref{H-g2-Stark})
considered in the present work.

Besides application for CQM~\cite%
{Garbe2020,Garbe2021-Metrology,Ilias2022-Metrology,Ying2022-Metrology} the
QFI also corresponds to the susceptibility of the fidelity\cite%
{Gu-FidelityQPT-2010,You-FidelityQPT-2007,You-FidelityQPT-2015} and the
appearance of a QFI peak signals a QPT in fidelity theory~\cite%
{Zhou-FidelityQPT-2008,Gu-FidelityQPT-2010,You-FidelityQPT-2007,You-FidelityQPT-2015,Zanardi-FidelityQPT-2006}%
, as applied to identify the frequency dependence of the QPT in the QRM~\cite{Ying-gC-by-QFI-2024}.

In this work we consider the coupling strength $g_{2}$ as the measurement
parameter $\lambda =g_{2}$. The wave wave function and QFI are calculated by
the exact diagonalization (ED) \cite%
{Ying2020-nonlinear-bias,Ying-Stark-top,Ying-g1g2hz-QFI-2025} and also
analyzed by the variational method in polaron picture\cite%
{Ying2015,Ying-gC-by-QFI-2024,Ying-g2hz-QFI-2024,Ying-g1g2hz-QFI-2025}.

\subsection{Globalized high measurement precision}

With the tunable critical point we can get globally high measurement
precision. Indeed, as shown in Fig.\ref{fig-Fq}(a), in the absence of the
Stark coupling (blue) the QFI is large around $g_{\mathrm{T}}$ ($\overline{g}%
_{2}=1)$ but becomes small away from $g_{\mathrm{T}}$. Now in the presence
of the Stark coupling, diverging QFI can be also available away from $g_{%
\mathrm{T}}$ as illustrated by some values of $\chi $ in Fig.\ref{fig-Fq}%
(a). Indeed, one can get large QFI continuously by tuning $\chi $, as
indicated by the 3D plot of $F_{Q}$ in Fig.\ref{fig-Fq}(b), and adding $\chi
$ seems to even enhance the QFI at fixed $\Omega =1.0\omega $. In fact, at
equi-$\widetilde{\Omega }_{c}$ lines in Fig.\ref{fig-Fq}(c), where
\begin{equation}
\widetilde{\Omega }_{c}=\Omega _{\chi }\sqrt{1-\chi ^{2}},\qquad \Omega
_{c}=\chi \omega +\frac{\widetilde{\Omega }_{c}}{\sqrt{1-\chi ^{2}}},
\end{equation}%
has a same value of spin flipping strength $\widetilde{\Omega }$ at $%
\overline{g}_{2c}^{\chi }$, the QFI has a similar diverging order as shown
in Fig.\ref{fig-Fq}(d). Actually later on we will see that the QFI manifests
more universal critical behavior. The globally available diverging QFI
promises a globalized high measurement precision for quantum metrology.

\begin{figure}[t]
\centering
\includegraphics[width=1.0\columnwidth]{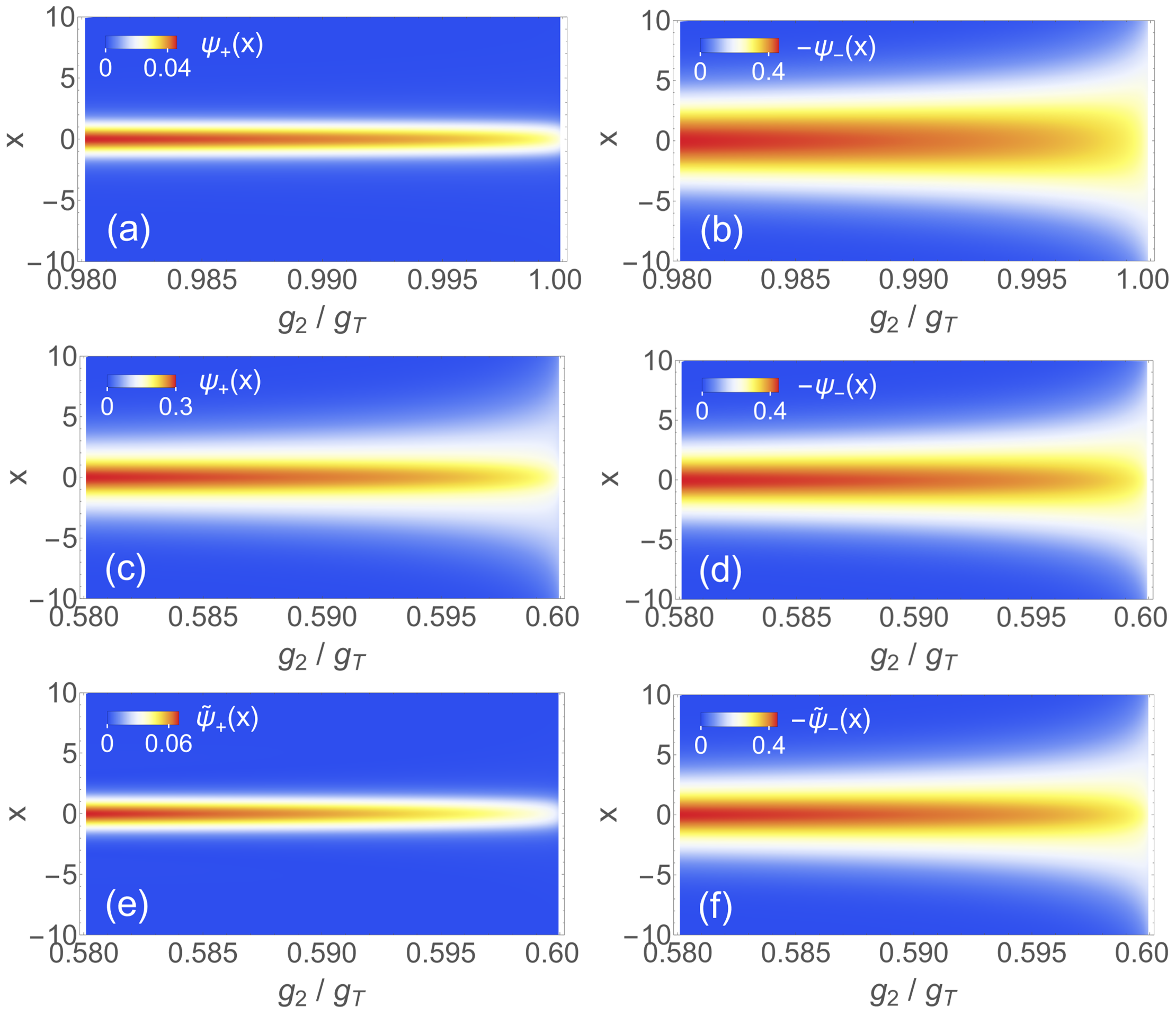}
\caption{Evolution of wave function with respect to $g_2$.
(a,b) $\psi _{\pm}(x)$ at $\chi=0$.
(c,d) $\psi _{\pm}(x)$ on unrotated spin basis at $\chi=0.8$.
(e,f) $\tilde\psi _{\pm}(x)$ on rotated spin basis at $\chi=0.8$.
$\tilde\Omega_c=0.1\omega$ in all panels ($\Omega=1.0\omega$ for $\chi=0.0$ and $\Omega=0.97\omega$ for $\chi=0.8$).}
\label{fig-WaveF}
\end{figure}
\begin{figure}[t]
\centering
\includegraphics[width=1.0\columnwidth]{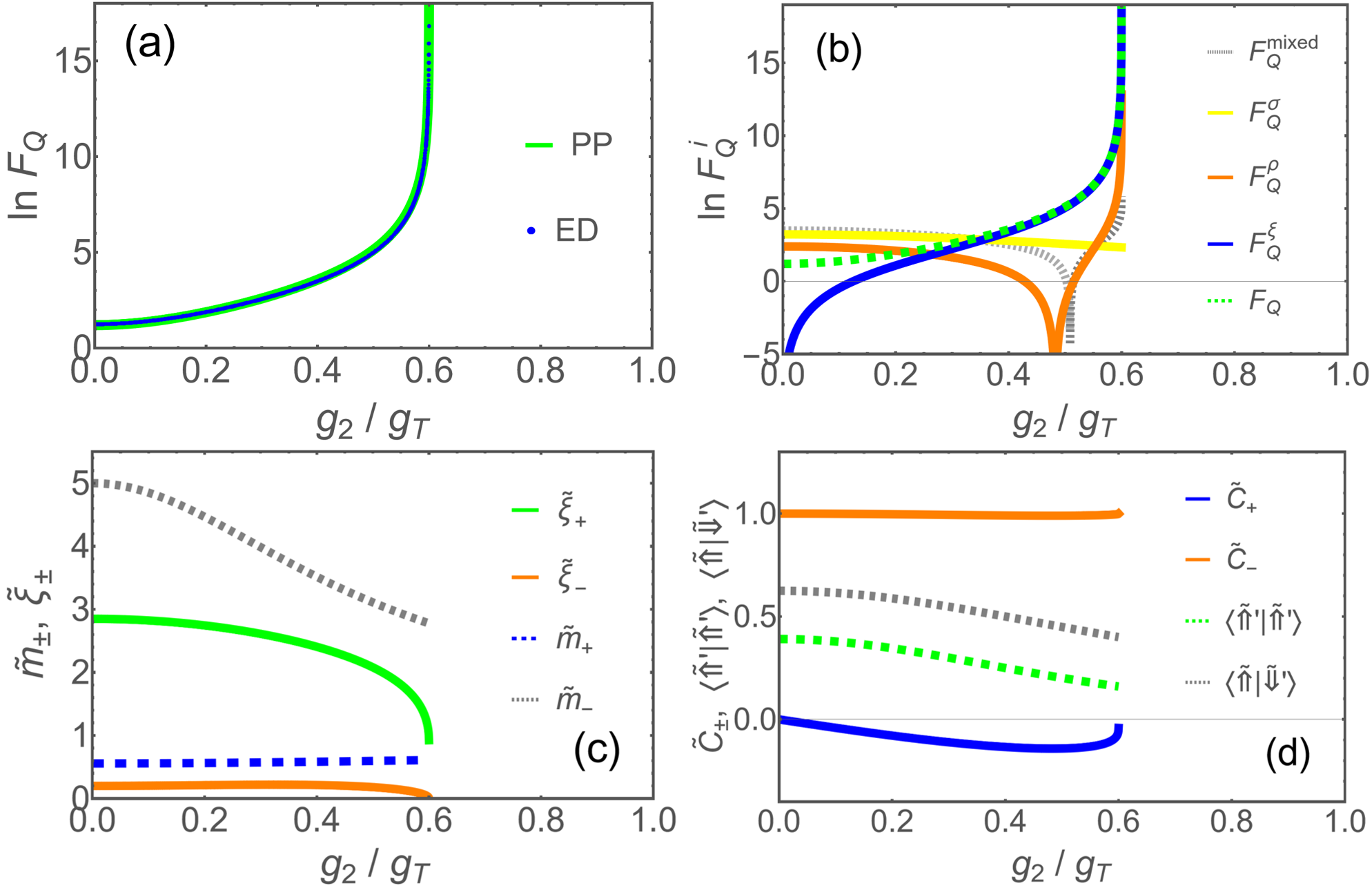}
\caption{Analysis of sensitivity resources:
(a) $F_Q$ by polaron picture (PP, green solid) in agreement with exact diagonalization (ED, blue dots).
(b) Contributions of squeezing ($F^{\xi}$, blue solid), basis weight ($F_Q^{\rho}$, orange solid), spin basis rotation ($F_Q^{\sigma}$, yellow solid) and mixed term ($F_Q^{\rm mixed}$, gray dotted) to the total $F_Q$ (green dashed). $F_Q^{\rm mixed}$ is negative before the sharp dip and $ln |F_Q^{\rm mixed}|$ is plotted.
(c) Evolutions of effective mass $\tilde{m} _{\pm}$ and basis frequency $\tilde{\xi} _{\pm}$.
(d) Evolutions of spin basis weight $\tilde{C} _{\pm}$ and spin basis variation products $\langle\tilde{\Uparrow}'|\tilde{\Uparrow}'\rangle$ ($ = \langle\tilde{\Downarrow}'|\tilde{\Downarrow}'\rangle$) and $\langle\tilde{\Uparrow}|\tilde{\Downarrow}'\rangle $ ($= -\langle\tilde{\Downarrow}|\tilde{\Uparrow}'\rangle$). In all panels $\chi=0.8$ and $\tilde\Omega _c=0.2$ ($\Omega =1.13\omega$).}
\label{fig-Fq-parts}
\end{figure}

\subsection{Sensitivity resources analyzed in polaron picture}

\label{Sect-resource-tracking}

It would be helpful to clarify the sensitivity resources by tracking the
contributions to the diverging QFI from different variations. Although the
eigenstate can be denoted either on unrotated spin basis or rotated spin
basis
\begin{equation}
\left\vert \psi \left( \lambda \right) \right\rangle =\psi _{+}|\Uparrow
\rangle +\psi _{-}|\Downarrow \rangle =\widetilde{\psi }_{+}|\widetilde{%
\Uparrow }\rangle +\widetilde{\psi }_{-}|\widetilde{\Downarrow }\rangle ,
\end{equation}%
the wave function $\widetilde{\psi }_{\pm }$ on the rotated spin basis is
easier to analyze. Fig.\ref{fig-WaveF} shows the evolutions of $\psi _{\pm }$
and $\widetilde{\psi }_{\pm }$ with respect to $g_{2}$, as calculated by ED
\cite{Ying2020-nonlinear-bias,Ying-Stark-top}. We see around the critical
point $g_{\mathrm{T}}$ in $\chi =0$ case,  $\psi _{+}$ becomes narrow [Figs.%
\ref{fig-WaveF}(a) while $\psi _{-}$ is extremely broadened \ref{fig-WaveF}%
(b)],. In $\chi \neq 0$ case, on the unrotated spin basis both $\psi _{+}$
and $\psi _{-}$ are broadened [Figs.\ref{fig-WaveF}(c) and \ref{fig-WaveF}%
(d)], while on the rotated basis $\widetilde{\psi }_{\pm }$  [Figs.\ref%
{fig-WaveF}(c) and \ref{fig-WaveF}(d)] have recover the narrowing and
broadening behavior as in $\chi =0$ case. Such a difference leads to a large
amplitude of $\langle \hat{\sigma }_{x}\rangle $ on the unrotated spin basis
[orange (light gray) solid line in Fig.\ref{fig-x2-p2-spinxz}(a)] but small $%
\langle \hat{\widetilde{\sigma }}_{x}\rangle $ on the rotated spin basis
[orange (light gray) solid line in Fig.\ref{fig-x2-p2-spinxz}(b)]. The
latter provides more convenience for the following resource tracking.

To facilitate the tracking of the contributions of different resources to
the QFI, we can apply the variational method in polaron picture\cite%
{Ying2015,Ying-gC-by-QFI-2024,Ying-g2hz-QFI-2024,Ying-g1g2hz-QFI-2025}. On
the rotated spin basis, the effective strength of spin flipping $\langle
\hat{\widetilde{\sigma }}_{x}\rangle $ tends to vanish around the critical
point due to small wave-packet overlap between narrow $\widetilde{\psi }_{+}$
and extremely broadened $\widetilde{\psi }_{-}$. In this situation, the two
rotated spin components are little mixed and one-polaron approximation can
capture well the physics. So we assume $\widetilde{\psi }_{\pm }=\widetilde{C%
}_{\pm }\widetilde{\varphi }_{\pm }$ with%
\begin{equation}
\widetilde{\varphi }_{\pm }=\frac{\left( \widetilde{m}_{\pm }\widetilde{\xi }%
_{\pm }\right) ^{1/4}}{\pi ^{1/4}}\exp \left[ -\frac{1}{2}\widetilde{m}_{\pm
}\widetilde{\xi }_{\pm }x^{2}\right]   \label{wave-PP}
\end{equation}%
representing the one polaron in variational polaron picture. The expression
of $\widetilde{C}_{\pm }$ and the variational method is presented in
Appendix \ref{Apendix-Wave-function}. Then, the QFI can be decomposed into
four parts%
\begin{equation}
F_{Q}=F_{Q}^{\rho }+F_{Q}^{\xi }+F_{Q}^{\sigma }+F_{Q}^{\mathrm{mixed}}\label{FQ-polaron}
\end{equation}%
where%
\begin{eqnarray}
F_{Q}^{\rho } &=&4\widetilde{C}_{+}^{\prime }\widetilde{C}_{+}^{\prime }+4%
\widetilde{C}_{-}^{\prime }\widetilde{C}_{-}^{\prime }, \\
F_{Q}^{\xi } &=&4\widetilde{C}_{+}\widetilde{C}_{+}\langle \widetilde{%
\varphi }_{+}^{\prime }|\widetilde{\varphi }_{+}^{\prime }\rangle +4%
\widetilde{C}_{-}\widetilde{C}_{-}\langle \widetilde{\varphi }_{-}^{\prime }|%
\widetilde{\varphi }_{-}^{\prime }\rangle , \\
F_{Q}^{\sigma } &=&4\widetilde{C}_{+}^{2}\langle \widetilde{\Uparrow }%
^{\prime }|\widetilde{\Uparrow }^{\prime }\rangle +4\widetilde{C}%
_{-}^{2}\langle \widetilde{\Downarrow }^{\prime }|\widetilde{\Downarrow }%
^{\prime }\rangle ,  \label{Fq-Spin}
\end{eqnarray}%
are respectively coming from variations of the wave packet ($F_{Q}^{\xi }$),
spin weight ($F_{Q}^{\rho }$), and rotation of spin basis ($F_{Q}^{\sigma }$%
), as indicated by the prime symbols. The mixed term%
\begin{eqnarray}
F_{Q}^{\mathrm{mixed}} &=&8\sum_{\sigma =\pm }\left( \widetilde{C}_{\sigma }%
\widetilde{C}_{\overline{\sigma }}^{\prime }\langle \widetilde{\varphi }%
_{\sigma }|\widetilde{\varphi }_{\overline{\sigma }}\rangle +\widetilde{C}%
_{\sigma }\widetilde{C}_{\overline{\sigma }}\langle \widetilde{\varphi }%
_{\sigma }|\widetilde{\varphi }_{\overline{\sigma }}^{\prime }\rangle
\right) \langle \overline{\sigma }|\sigma ^{\prime }\rangle   \nonumber \\
&=&8\langle \widetilde{\Downarrow }|\widetilde{\Uparrow }^{\prime }\rangle
\lbrack (\widetilde{C}_{-}\widetilde{C}_{+}^{\prime }-\widetilde{C}_{+}%
\widetilde{C}_{-}^{\prime })\langle \widetilde{\varphi }_{-}|\widetilde{%
\varphi }_{+}\rangle   \nonumber \\
&&+8\widetilde{C}_{-}\widetilde{C}_{+}(\langle \widetilde{\varphi }_{-}|%
\widetilde{\varphi }_{+}^{\prime }\rangle -\langle \widetilde{\varphi }_{+}|%
\widetilde{\varphi }_{-}^{\prime }\rangle )]  \label{Fq-mixed}
\end{eqnarray}%
with $\overline{\sigma }=-\sigma $, arises from the non-vanishing products ($%
\langle \overline{\sigma }|\sigma ^{\prime }\rangle $) of the rotated spin
basis. We provide the explicit expressions of $\langle \widetilde{\Downarrow
}^{\prime }|\widetilde{\Downarrow }^{\prime }\rangle $, $\langle \widetilde{%
\Uparrow }^{\prime }|\widetilde{\Uparrow }^{\prime }\rangle $ and $\langle
\widetilde{\Downarrow }^{\prime }|\widetilde{\Uparrow }\rangle $, $\langle
\widetilde{\Downarrow }|\widetilde{\Uparrow }^{\prime }\rangle $ in Appendix %
\ref{Appendix-Spin-Rotation}.

In Fig.\ref{fig-Fq-parts}(a) we compare $F_{Q}$ by the polaron picture on (%
\ref{wave-PP}) [green solid line] which agrees well with the result of ED
(blue dots). Fig.\ref{fig-Fq-parts}(b) tracks the decomposed contributions
from $F_{Q}^{\xi }$ [blue (dark gray) solid], $F_{Q}^{\rho }$ [orange (gray)
solid], $F_{Q}^{\sigma }$ [yellow (light gray) solid], $F_{Q}^{\mathrm{mixed}%
}$ (gray dotted) are tracked in Fig.\ref{fig-Fq-parts}(b), we see that
although $F_{Q}^{\xi }$ $F_{Q}^{\rho }$ and $F_{Q}^{\mathrm{mixed}}$ are all
singular-like, $F_{Q}^{\xi }$ is the leading contribution to the total $F_{Q}
$ [green dashed].

In fact, the leading part $F_{Q}^{\xi }$ dominantly stems from squeezing as
the frequency [$\widetilde{\varpi }_{\pm }$, solid lines in Fig.\ref%
{fig-Fq-parts}(c)] variations are singular-like around the critical point
while the mass variations are not [$\widetilde{m}_{\pm }$, broken lines in
Fig.\ref{fig-Fq-parts}(c)]. Other contributions are smaller: The $F_{Q}^{%
\mathrm{mixed}}$ is smaller due to small overlap $\langle \widetilde{\varphi
}_{-}|\widetilde{\varphi }_{+}\rangle $ as in afore-mentioned behavior of $%
\langle \hat{\widetilde{\sigma }}_{x}\rangle $ and small product $\widetilde{%
C}_{-}\widetilde{C}_{+}$ due to the unbalanced spin-component weight [solid
lines in Fig.\ref{fig-Fq-parts}(d)], $F_{Q}^{\sigma }$ is flat as $\langle
\widetilde{\Uparrow }^{\prime }|\widetilde{\Uparrow }^{\prime }\rangle
=\langle \widetilde{\Downarrow }^{\prime }|\widetilde{\Downarrow }^{\prime
}\rangle $ is small [orange dashed line in Fig.\ref{fig-Fq-parts}(d)]. The $%
F_{Q}^{\rho }$ is also relatively smaller than $F_{Q}^{\xi }$ as the weight
variations of $\widetilde{C}_{\pm }$ comes indirectly from squeezing. As a
result, the leading contribution comes from the resource of squeezing.

\begin{figure}[t]
\centering
\includegraphics[width=0.68\columnwidth]{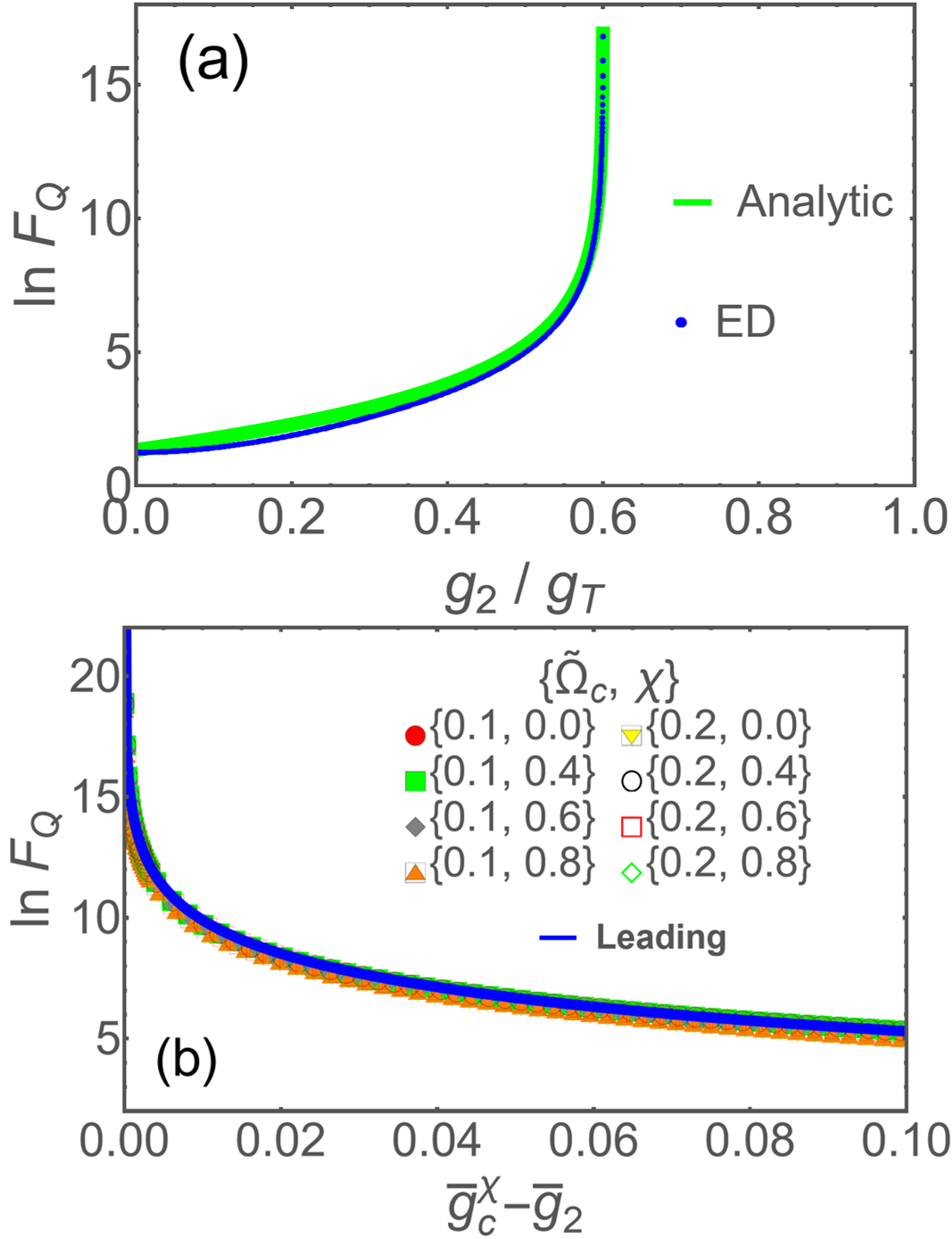}
\caption{Critical exponent and universality of QFI. (a) Comparison of major-order expression of $F_Q$ [Eq.~(\ref{Fq-main-orders}), green solid line] with ED result (blue dots). (b) Universal $F_Q$ by ED (symbols) with different values of $\tilde\Omega$ and $\chi$ around the critical point, coinciding with the critical exponent $\gamma =2$ in Eq.~(\ref{Fq-exponent}) from the leading term (blue solid).  }
\label{fig-Fq-Exponent}
\end{figure}

\subsection{Critical exponent and universality of quantum Fisher information}
\label{Sect-exponent}

We further extract the critical exponent analytically and reveal the
universality in the divergence of the QFI. In the vicinity of the critical
point, by expansion on the distance from the critical point (Appendix \ref{Apendix-Wave-function}) we obtain the
major orders of the QFI as
\begin{equation}
F_{Q}\approx \frac{\left( \overline{g}_{c}^{\chi }-\overline{g}_{2}\right)
^{-2}}{8g_{\mathrm{T}}^{2}}-\frac{\left[ \chi ^{2}-w_{z}\left( 2-\chi
^{2}\right) \right] \left( \overline{g}_{c}^{\chi }-\overline{g}_{2}\right)
^{-1}}{16\left( 1+w_{z}\right) \sqrt{1-\chi ^{2}}g_{\mathrm{T}}^{2}}.
\label{Fq-main-orders}
\end{equation}%
Eq. (\ref{Fq-main-orders}) [green line in Fig. \ref{fig-Fq-Exponent}(a)]
well captures the critical behavior of QFI, as compared to the ED result
[blue dots in Fig. \ref{fig-Fq-Exponent}(a)].

We see in the leading critical behavior,
\begin{equation}
F_{Q}\sim \frac{1}{8g_{\mathrm{T}}^{2}}\left( \overline{g}_{c}^{\chi }-%
\overline{g}_{2}\right) ^{-\gamma }  \label{Fq-exponent}
\end{equation}%
which is universal for different values of $\chi $, not only with the same
critical exponent
\begin{equation}
\gamma =2
\end{equation}%
but also with the same coefficient $1/8g_{\mathrm{T}}^{2}$. Such
universality guarantees a same order of precision available when the
critical point is tuned, as demonstrated by Fig. \ref{fig-Fq-Exponent}(b). A
robust universality of QFI was noticed in the non-Hermitian JC mode\cite%
{Ying-Topo-JC-nonHermitian-Fisher}, while here we have the universality of
QFI in Hermitian case.

\section{Globally strong squeezing resource}

\label{Section-Global-Squeezing}

As mentioned in Introduction, squeezing as a quantum resource is robust
against decoherence and dissipation~\cite{Gietka2023PRL-Squeezing,LiJieYouJQ2022SqueezingAgaistTemperature,WengYouJQ2025SqueezingRobust,Buzek1992SqueezingAgainstdissipation}. Here we have a globally
strong squeezing resource. Indeed the leading sensitivity resource (\ref%
{Fq-exponent}) comes from the renormalization of the frequency which is a
reflection of squeezing. To see the squeezing more clearly we can check the
Wigner function~\cite%
{Wigner1932,WignerReview2018,Ying-gapped-top,Ying-g1g2hz-QFI-2025}
\begin{equation}
W_{\pm }\left( x,p\right) =\frac{1}{2\pi }\int\limits_{-\infty }^{\infty
}e^{ipy}\widetilde{\psi }_{\pm }^{\ast }(x+\frac{y}{2})\widetilde{\psi }%
_{\pm }(x-\frac{y}{2})dy,
\end{equation}%
where we have set $\hbar =1$. Without squeezing the profile of Wigner
function $W_{\pm }\left( x,p\right) $ is round in the $x$-$p$ plane, while
the flattening of $W_{\pm }\left( x,p\right) $ signals the presence of
squeezing, with a more flattened $W_{\pm }\left( x,p\right) $ meaning a
stronger squeezing.

\begin{figure}[t]
\centering
\includegraphics[width=1.0\columnwidth]{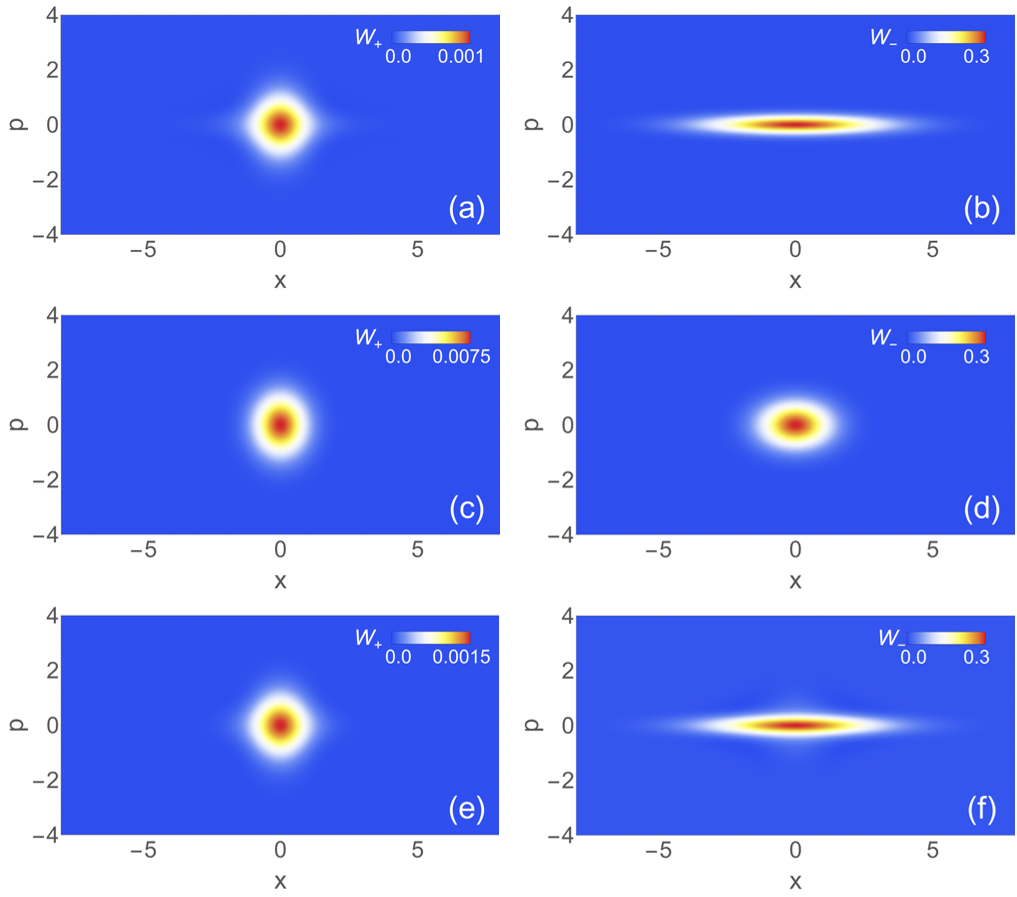}
\caption{Wigner function $W_{\pm}$ and globalized strong squeezing resource.
(a,b) Strong squeezing around $\overline{g}_2=1.0$ for $\chi=0.0$ ($\overline{g}_2=0.99$ here) .
(c,d) Weakened squeezing away from $\overline{g}_2=1$ for $\chi=0.0$  ($\overline{g}_2=0.59$ illustrated here).
(e,f) Regained strong squeezing by introducing $\chi$ ($\overline{g}_2=0.594$, $\chi=0.8$ here).
We illustrate by $\tilde\Omega_c=0.1\omega$ in all panels.}
\label{fig-Wigner}
\end{figure}

We show $W_{\pm }\left( x,p\right) $ in Fig.\ref{fig-Wigner} for the ground
state. In the absence of the Stark coupling, as in Figs.\ref{fig-Wigner}(a)
and \ref{fig-Wigner}(b), we see a strong squeezing effect especially in the
spin-down component around the original critical point $g_{\mathrm{T}}$.
However the squeezing is weakened away from $g_{\mathrm{T}}$, as illustrated
in Figs.\ref{fig-Wigner}(c) and \ref{fig-Wigner}(d). Nevertheless, the
strong squeezing is restored in the presence of the Stark coupling as
demonstrated in Figs.\ref{fig-Wigner}(e) and \ref{fig-Wigner}(f). Thus, we
can always have a strong squeezing resource in different values of coupling.
Note that this squeezing is the leading quantum resource for the critical
QFI, as addressed in Sections \ref{Sect-resource-tracking} and \ref%
{Sect-exponent}, we indeed have a protocol of global nonlinear CQM with always strong squeezing resource.

\begin{figure}[t]
\centering
\includegraphics[width=1.0\columnwidth]{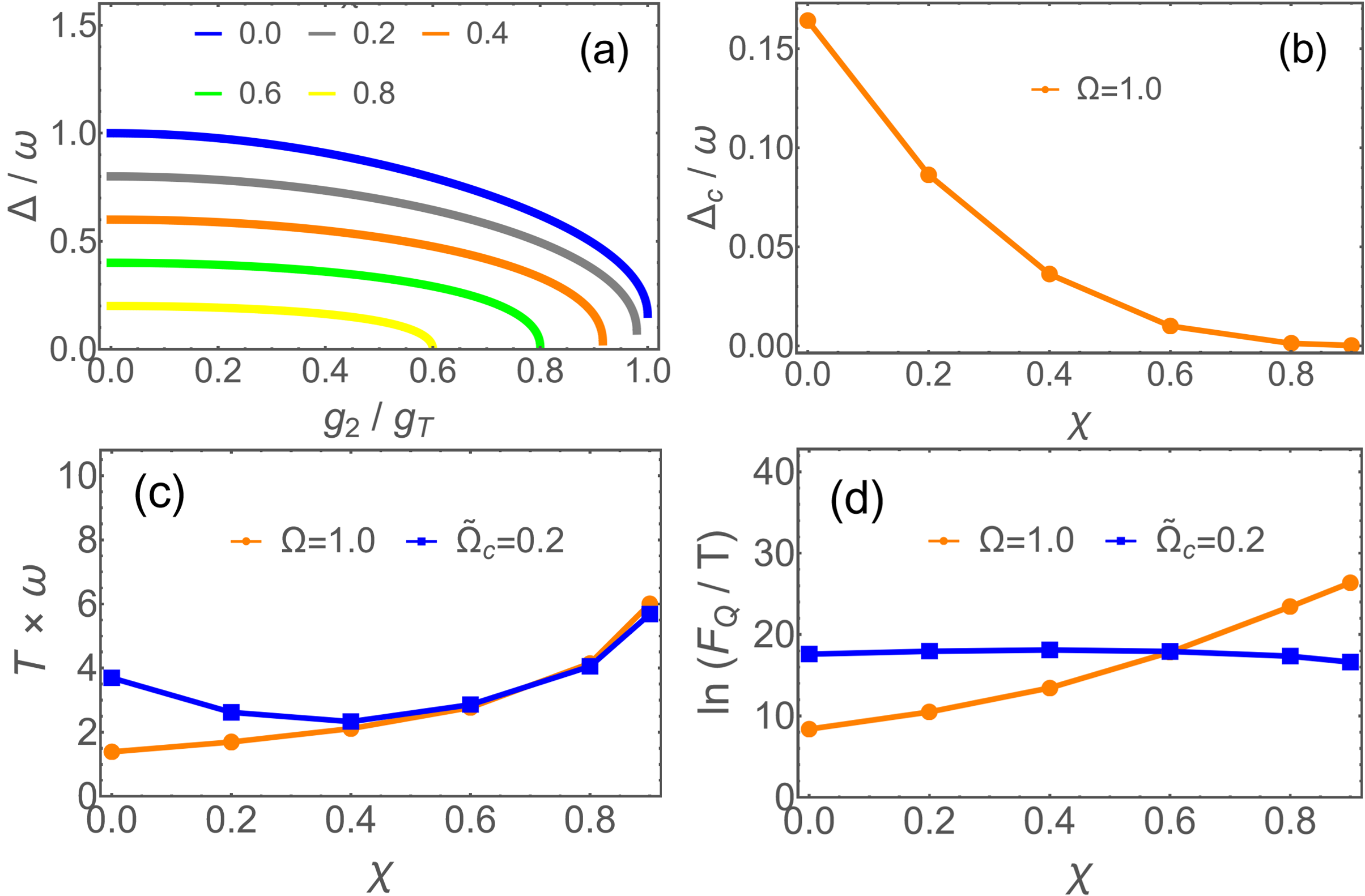}
\caption{Finite preparation time of probe state.
(a) Gap evolution with respect to $g_2$ at different values of $\chi$.
(b) Finite gap at the critical points $\overline{g}_c^{\chi}$.
(c) Finite preparation time of probe state $T$.
(d) Protocol applicability assessed by $F_Q/T$ at $\overline{g}_c^{\chi}$.
$\Omega=1.0\omega$ in (a) and (b), while $\Omega=1.0\omega$ (orange dots) and $\tilde\Omega _c=0.2\omega$ (blue squares) in (c) and (d). }
\label{fig-gap-time}
\end{figure}

\section{Finite preparation time for probe state (PTPS)}

\label{Section-Finite-T}

Another merit of nonlinear CQM is finite PTPS\cite%
{Ying2022-Metrology,Ying-g2hz-QFI-2024,Ying-g1g2hz-QFI-2025}, while the PTPS
is diverging in the linear CQM based on the quantum
phase transition of the linear QRM\cite{Garbe2020,Ying2022-Metrology}. Such
a finite-PTPS merit remains in our protocol. In adiabatic preparation of the
probe state the PTPS is inversely proportional to the gap $\Delta $, so that
the PTPS can be estimated by~\cite{Garbe2020,Ying2022-Metrology}
\begin{equation}
T=\int\limits_{0}^{\overline{g}_{c}^{\chi }}\Delta \left( \overline{g}%
_{2}\right) ^{-1}d\overline{g}_{2}
\end{equation}%
where we have swept from zero quadratic coupling fully to the critical
point $\overline{g}_{c}^{\chi }$ in the presence of the Stark coupling [Eq.(%
\ref{gC-at-Stark})].

The gap evolution is shown in Fig.\ref{fig-gap-time}(a) at various values of
$\chi $ with fixed $\Omega =1.0\omega $. The gap minima at $\overline{g}%
_{c}^{\chi }$ are collected in Fig.\ref{fig-gap-time}(b), remaining always
nonvanishing despite the decreasing tendency with respect to $\chi $. Such a
finite gap is determined by
\begin{equation}
\Delta = 2\sqrt{e_{-}^{2}+\left( \widetilde{S}_{\Omega }+\widetilde{S}_{\kappa }\right) ^{2}},
\end{equation}
as derived in Appendix \ref{Apendix-Wave-function}, which has non-vanishing minimum
$\widetilde{S}_{\Omega }+\widetilde{S}_{\kappa }$.

The final PTPS is plotted in Fig.\ref{fig-gap-time}(c) for $\Omega
=1.0\omega $ (orange dots) and $\widetilde{\Omega }_{c}=0.1\omega $\ (blue
squares), basically remaining in a finite order. One can assess the
practicability of the protocol by the ratio of the QFI and PTPS\cite%
{Ying2022-Metrology}, which is shown in Fig.\ref{fig-gap-time}(d) and
showing similar orders of practicability at different Stark couplings.

\section{Conclusions}

\label{Section-Conclusion}

We have proposed to combine the Stark coupling with the conventional
quadratic coupling to globalize the nonlinear CQM by
the two-photon Rabi-Stark model in light-matter interaction. The
introduction of the Stark coupling extend the single critical point $g_{%
\mathrm{T}}$ of the two-photon QRM to a continuously tunable critical point $%
g_{\mathrm{T}}^{\chi }$, thus the local restriction of the critical behavior
can be broken and the criticality is available in a wide range of coupling
for application of CQM.

Indeed, by tuning the Stark coupling the QFI can be divergently high not
only around $g_{\mathrm{T}}$ of the two-photon QRM but also in the entire
range of coupling from zero to $g_{\mathrm{T}}$. By analysis on the
decomposed contributions to the QFI, we have shown that the leading
contribution to the critical QFI is coming from squeezing. We have
analytically extracted the critical exponent of the QFI to be $\gamma =2$.
Moreover, the critical QFI turns out to be universal at different Stark
couplings and flipping strengths. The criticality and universality of
QFI guarantees a similar order of measurement precision in quantum metrology.

By analysis on the Wigner function we have confirmed that a strong squeezing
is globally available, which is favorable in practice as squeezing as a
quantum resource is known to be robust against decoherence and dissipation~\cite{Gietka2023PRL-Squeezing,LiJieYouJQ2022SqueezingAgaistTemperature,WengYouJQ2025SqueezingRobust,Buzek1992SqueezingAgainstdissipation}.
Our protocol also retains the merit that the PTPS is finite. Here, we have
not stressed for the other merit that our protocol is applicable to finite
frequencies, which actually has been illustrated in Figs.~\ref{fig-Fq}, \ref%
{fig-Fq-parts}, and \ref{fig-Fq-Exponent} where the frequency is not restricted to the low-frequency limit as in
the linear QRM to gain the criticality.

In short, we have proposed to exploit the two-photon Rabi-Stark model to
realize a globalized nonlinear CQM, with the
advantages of globally high measurement precision, globally strong
squeezing, applicability to finite frequencies and avoidance of detrimental
problem of diverging PTPS.

As a final remark, the two-photon Rabi-Stark model can be realized in
trapped ions and superconducting circuits\cite%
{Felicetti2018-mixed-TPP-SPP,Stark-Grimsmo2013,Stark-Cong2020,Zhai2025TwoPhotonStark}%
, thus our protocol is practically relevant. It should also be mentioned
that the system is unstable beyond the critical point due to spectral
collapse\cite%
{Felicetti2018-mixed-TPP-SPP,Felicetti2015-TwoPhotonProcess,e-collpase-Garbe-2017,Rico2020,e-collpase-Duan-2016,CongLei2019,Ying-Stark-top}%
, which, nevertheless, can be resolved and stabilized by adding a realizable
quartic term\cite{Ying2025g2A4,HangHang2024A4}. Such a situation deserves a
special discussion which we shall address elsewhere.

\section*{Acknowledgment}

This work was supported by the National Natural Science Foundation of China
(Grants No. 12474358, No. 11974151, and No. 12247101).

\appendix\bigskip

\section{Rotated spin basis in the presence of the Stark coupling}

\label{Appendix-Spin-Rotation}

In the presence of the Stark coupling, corresponding to the rotated spin (%
\ref{Sxz-tilde}), the eigenstates of $\hat{\widetilde{\sigma }}_{z}$
\[
\hat{\widetilde{\sigma }}_{z}|\widetilde{\Uparrow }\rangle =+|\widetilde{%
\Uparrow }\rangle ,\qquad \hat{\widetilde{\sigma }}_{z}|\widetilde{%
\Downarrow }\rangle =-|\widetilde{\Downarrow }\rangle ,
\]
take the form%
\begin{eqnarray}
|\widetilde{\Uparrow }\rangle  &=&\frac{1}{\sqrt{2}}\sqrt{1+\cos \vartheta }%
|\Uparrow \rangle +\frac{sign[\chi ]}{\sqrt{2}}\sqrt{1-\cos \vartheta }%
|\Downarrow \rangle , \\
|\widetilde{\Downarrow }\rangle  &=&-\frac{sign[\chi ]}{\sqrt{2}}\sqrt{%
1-\cos \vartheta }|\Uparrow \rangle +\frac{1}{\sqrt{2}}\sqrt{1+\cos
\vartheta }|\Downarrow \rangle .
\end{eqnarray}%
And reversely we have the unrotated spin basis in terms of rotated ones%
\begin{eqnarray}
| &\Uparrow &\rangle =\frac{1}{\sqrt{2}}\sqrt{1+\cos \vartheta }|\widetilde{%
\Uparrow }\rangle -\frac{sign[\chi ]}{\sqrt{2}}\sqrt{1-\cos \vartheta }|%
\widetilde{\Downarrow }\rangle , \\
| &\Downarrow &\rangle =\frac{sign[\chi ]}{\sqrt{2}}\sqrt{1-\cos \vartheta }|%
\widetilde{\Uparrow }\rangle +\frac{1}{\sqrt{2}}\sqrt{1+\cos \vartheta }|%
\widetilde{\Downarrow }\rangle .
\end{eqnarray}%
Here the rotation angle of the spin and the derivative with respect to the
coupling are given by%
\begin{equation}
\cos \vartheta =\frac{\overline{g}_{2}}{\sqrt{\overline{g}_{2}^{2}+\chi ^{2}}%
},\quad \frac{d\cos \vartheta }{dg_{2}}=\frac{\chi ^{2}}{\left( \overline{g}%
_{2}^{2}+\chi ^{2}\right) ^{3/2}}\frac{1}{g_{\mathrm{T}}}.
\end{equation}%
The products of spin basis in the variation of $g_{2}$ are then direct to
obtain
\begin{equation}
\langle \widetilde{\Uparrow }^{\prime }|\widetilde{\Uparrow }^{\prime
}\rangle =\frac{1}{8}\left( \frac{d\cos \vartheta }{dg_{2}}\right)
^{2}\left( \frac{1}{1+\cos \vartheta }+\frac{1}{1-\cos \vartheta }\right)
\end{equation}%
and $\langle \widetilde{\Downarrow }^{\prime }|\widetilde{\Downarrow }%
^{\prime }\rangle =\langle \widetilde{\Uparrow }^{\prime }|\widetilde{%
\Uparrow }^{\prime }\rangle $ in (\ref{Fq-Spin}), while%
\begin{equation}
\langle \widetilde{\Downarrow }|\widetilde{\Uparrow }^{\prime }\rangle =%
\frac{sign[\chi ]}{4}\frac{d\cos \vartheta }{dg_{2}}\left( \frac{\sqrt{%
1+\cos \vartheta }}{\sqrt{1-\cos \vartheta }}+\frac{\sqrt{1-\cos \vartheta }%
}{\sqrt{1+\cos \vartheta }}\right)
\end{equation}%
and $\langle \widetilde{\Downarrow }^{\prime }|\widetilde{\Uparrow }\rangle
=-\langle \widetilde{\Downarrow }|\widetilde{\Uparrow }^{\prime }\rangle $
in (\ref{Fq-mixed}).

\section{Variational method for tracking sensitivity resources in QFI}
\label{Apendix-Wave-function}

With the variational wave function $\widetilde{\psi }_{\pm }=\widetilde{C}%
_{\pm }\widetilde{\varphi }_{\pm }$ where $\widetilde{\varphi }_{\pm }$ is
given in Eq.(\ref{wave-PP}), the Hamiltonian can be represented in matrix
form
\begin{equation}
H=\left(
\begin{array}{cc}
\widetilde{\varepsilon }_{+} & \widetilde{S}_{\Omega }+\widetilde{S}_{\kappa
} \\
\widetilde{S}_{\Omega }+\widetilde{S}_{\kappa } & \widetilde{\varepsilon }%
_{-}%
\end{array}%
\right)
\end{equation}%
in the subspace of the lowest energies, with the single-particle energy in
the diagonal terms
\begin{equation}
\widetilde{\varepsilon }_{\pm }=\frac{\widetilde{\xi }_{\pm }\omega }{4}\pm
\widetilde{\epsilon }+\left( 1\pm \sqrt{\overline{g}_{2}^{2}+\chi ^{2}}%
\right) \frac{\omega }{4\widetilde{m}_{\pm }\widetilde{\xi }_{\pm }}
\nonumber
\end{equation}%
and the tunneling or spin-flipping energy $S_{\Omega }=\frac{\Omega }{2}%
\langle \varphi _{+}|\varphi _{-}\rangle $ in the non-diagonal terms,
explicitly
\begin{eqnarray}
\widetilde S_{\Omega } &=&\frac{(\widetilde{m}_{+}\widetilde{m}_{-}\widetilde{\xi }_{+}%
\widetilde{\xi }_{-})^{1/4}\widetilde{\Omega }}{\sqrt{2}(\widetilde{m}_{+}%
\widetilde{\xi }_{+}+\widetilde{m}_{-}\widetilde{\xi }_{-})^{1/2}},
\label{S-Omega} \\
\widetilde{S}_{\kappa } &=&\frac{\sqrt{2}(\widetilde{m}_{+}\widetilde{m}_{-}%
\widetilde{\xi }_{+}\widetilde{\xi }_{-})^{5/4}\widetilde{\kappa }}{(%
\widetilde{m}_{+}\widetilde{\xi }_{+}+\widetilde{m}_{-}\widetilde{\xi }%
_{-})^{3/2}}.
\end{eqnarray}%
$\widetilde{m}_{\pm }$, $\widetilde{\Omega }$ and $\widetilde{\kappa }$ are
given in Eqs. (\ref{mass-tilde}), (\ref{Omega-tilde}) and (\ref{kapa-tilde}%
). The eigen energy $E^{\eta }=e_{+}+\eta \sqrt{e_{-}^{2}+\left( \widetilde{S%
}_{\Omega }+\widetilde{S}_{\kappa }\right) ^{2}}$, where $e_{\pm }=(%
\widetilde{\varepsilon }_{+}\pm \widetilde{\varepsilon }_{-})/2$, has two
branches labelled by $\eta =\pm $, while the ground state takes $\eta =-$.
The spin-component weight are determined by
\begin{equation}
\widetilde{C}_{+}=B_{+}/\sqrt{B_{+}^{2}+B_{-}^{2}},\qquad \widetilde{C}%
_{-}=B_{-}/\sqrt{B_{+}^{2}+B_{-}^{2}},  \label{c1c2}
\end{equation}%
where%
\begin{eqnarray}
B_{+} &=&(e_{-}-\sqrt{e_{-}^{2}+\left( \widetilde{S}_{\Omega }+\widetilde{S}%
_{\kappa }\right) ^{2}}, \\
B_{-} &=&\left( \widetilde{S}_{\Omega }+\widetilde{S}_{\kappa }\right) ,
\end{eqnarray}%
The spin-component weights fulfill the normalization condition $\widetilde{C}%
_{+}^{2}+\widetilde{C}_{-}^{2}=1$.

The energy difference of the two energy branches
\begin{equation}
\Delta =E^{+}-E^{-}=2\sqrt{e_{-}^{2}+\left( \widetilde{S}_{\Omega }+%
\widetilde{S}_{\kappa }\right) ^{2}}  \label{gap-general}
\end{equation}%
will be the gap when there is no level crossing in the lowest excited states.

The wave-packet frequencies in the two spin component, $\widetilde{\xi }%
_{\pm }$, are variationally determined by numerical minimization of $E^{-}$,
with the result illustrated by the solid lines in Fig.\ref{fig-WaveF}(c).
Substitutions of $\widetilde{\xi }_{\pm }$ and $\widetilde{C}_{\pm }$ in Eq.(%
\ref{FQ-polaron}) gives the total QFI. The final QFI by the above
variational method gives the green solid line in Fig.\ref{fig-WaveF}(a), in
good agreements with the ED result (blue dots).

Setting $\widetilde{\xi }_{\pm }\approx \widetilde{\varpi }_{\pm }$
approximately and carrying out an expansion by small distance to the
critical point, $\overline{g}_{c}^{\chi }-\overline{g}_{2}$, we get the
leading orders analytically in Eq. (\ref{Fq-main-orders}).

\bibliography{Refs-2025}

\end{document}